# HIERARCHICAL FEATURES OF LARGE-SCALE CORTICAL CONNECTIVITY

L. da F. Costa[1] and O. Sporns[2]

1 – Instituto de Física de São Carlos,
   Universidade de São Paulo,
   Caixa Postal 369, São Carlos, SP
   13560-970, Brazil
   luciano@if.sc.usp.br

2 – Department of Psychology
   Indiana University
   Bloomington, IN 47405
   USA
   osporns@indiana.edu

ABSTRACT

The analysis of complex networks has revealed patterns of organization in a variety of natural and artificial systems, including neuronal networks of the brain at multiple scales. In this paper, we describe a novel analysis of the large-scale connectivity between regions of the mammalian cerebral cortex, utilizing a set of hierarchical measurements proposed recently. We examine previously identified functional clusters of brain regions in macaque visual cortex and cat cortex and find significant differences between such clusters in terms of several hierarchical measures, revealing differences in how these clusters are embedded in the overall cortical architecture. For example, the ventral cluster of visual cortex maintains structurally more segregated, less divergent connections than the dorsal cluster, which may point to functionally different roles of their constituent brain regions.

## 1. INTRODUCTION

The mammalian cerebral cortex is possibly one of the most complex systems found in nature, forming an intricate pattern of connections between individual neurons, specialized neuronal populations and cortical regions. Large-scale patterns of interregional corticocortical connections exhibit distinct patterns, characterized by clustering of functionally related brain regions, combined with short paths and wiring lengths. The resulting functional duality between localization/modularization and distribution/integration [1,2] creates dynamical states that underlie perception and cognition and can be accessed using modern invasive (neurophysiology) and noninvasive (EEG, fMRI) neuroscience methods. As an increasing amount of experimental data becomes available, it has become particularly interesting to represent, analyze and model such data in order to emphasize important organizational and functional properties of the mammalian cortical structure.

The analysis of complex networks (e.g. [3-6]) has recently been recognized as a powerful and flexible approach for representing, analyzing and modeling a broad range of natural and artificial systems. Indeed, complex networks can be thought of as an intersection between graph theory and statistical physics, thus incorporating and integrating several powerful and general concepts and methods from these two well-established areas. In neuroscience, complex networks have been proposed as models of microscopic and mesoscopic neuronal connectivity. Wiring morphology has been related to functional circuit properties such as synchronization [7] and morphologically realistic neuronal complex networks have been used in order to investigate the relationship between neuronal shape and connectivity (e.g. [8]), and to devise statistical models of neuron-to-neuron connectivity [9-11]. At the macroscopic or large scale, complex network tools have revealed clusters of functionally related areas [12], the presence of small-world attributes [13,14], high proportions of cycles and specific network motifs [15].

The connectivity of a complex network can be characterized in terms of several topological measurements. While many studies have considered node degree, clustering coefficient, and shortest paths between two nodes, such measurements provide rather limited, though important, information about network structure. For instance, there is an infinite number of networks which lead to identical mean values for these three measurements (see, for instance, [6]). Several measurements have been suggested in order to provide additional and richer characterization of the connectivity of complex networks [6]. Recently, a set of hierarchical measures was proposed [16-18] which explicitly take into account the hierarchical organization established around each node. For instance, the traditional node degree and clustering coefficients can naturally be extended as hierarchical signatures in terms of the hierarchical levels. Because such measures capture a much broader context around each node, they have potential for richer characterization of the local and global network connectivity. Preliminary applications of these hierarchical measures to Sznajd complex networks [19] as well as real data related to word associations and amino acids [16] have substantiated their strong discriminative power.

The existence of distinct clusters of areas in all large-scale cortical connectivity matrices examined so far [12] raises the question of whether these clusters differ in terms of their local and global connectivity, as well as in terms of hierarchical measures that evaluate their embedding in the overall architecture. In this paper we utilize an array of hierarchical measures to identify potential differences in previously reported clusters of cortical regions. Our set of hierarchical measures reveals differences in the way that individual brain regions access and interact with the remainder of the network. We find that, by providing richer information about the connectivity context around each region, the adopted hierarchical measurements can reveal relevant structural properties for distinct clusters of regions. Consistently with previous investigations showing differences between the organization of the dorsal and ventral macaque visual regions, our analyses revealed that the ventral system incorporates areas which tend to be more strongly connected at hierarchical levels higher than 1, while exhibiting less divergence than the dorsal system. When applied to cat cortical regions, the hierarchical measures revealed that sensory areas are functionally more segregated than those of the frontolimbic complex. Overall, the obtained results allowed the identification of new principles of cortical organization in different systems and species.

We begin by presenting and illustrating each of the four adopted hierarchical measurements and we then describe our experimental data sets. The results for macaque visual cortical areas and cat cortical areas are then presented and discussed, and we conclude with a discussion of the main findings and perspectives for further developments.

2. HIERARCHICAL MEASUREMENTS

A complex network $G$ is composed of a set of $N$ nodes and a set of $K$ edges between such nodes. If $i$ and $j$ are generic nodes of $G$; a directed edge extending from $i$ to $j$ is henceforth represented by the ordered pair $(i, j)$. Therefore, the maximum number of directed edges which can be established amongst $N$ nodes, excluding self-connections, is $T = N(N-1)$. The shortest path from a node $i$ to another node $j$ corresponds to the sequence of edges starting at $i$ and extending up to $j$ which involves the smallest number of edges, defining its length.

Given a specific node $i$, the set of nodes which are exactly at shortest distance $d$ from $i$ is henceforth called the ring of radius (or distance) $d$ centered at $i$, expressed as $R_d(i)$. Figure 1 illustrates a simple network with $N = 10$ nodes and $K = 15$ edges and the rings of radius $d = 1$ and 2 centered at node $i = 5$. Each value of $d$ therefore establishes a respective hierarchical level with respect to the reference node $i$. Note that different rings at the same reference radius $d$ are usually obtained for distinct reference nodes. Indeed, rings at successive distances around a reference node $i$ provide a hierarchical representation of the whole network with respect to $i$.

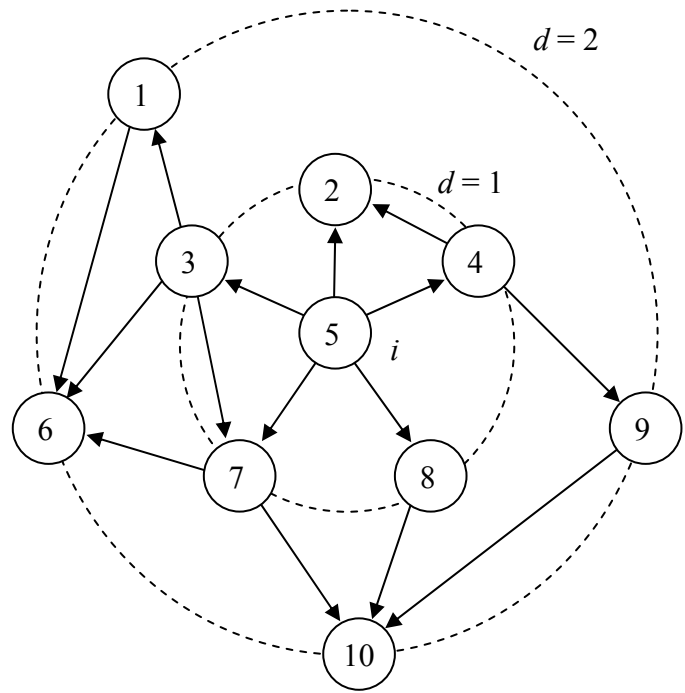

Figure 1 – A simple directed network containing $N = 10$ nodes and $K = 14$ edges. The rings of radiuses 1 and 2 centered at $i = 5$, i.e. $R_1(5)$ and $R_2(5)$, are identified by the two dashed circles.

The number of hierarchical neighbors at distance $d$ from a node $i$, hence represented as $n_d(i)$, is henceforth understood as the number of nodes contained in the ring $R_d(i)$. For example, in Figure 1, $n_1(5) = 5$ and $n_2(5) = 4$. Note that $n_d(i)$ goes to zero as the border of the network (i.e. the set of nodes with null outdegree) is reached. Note also that the sum of all hierarchical neighbors over all hierarchical depths must be equal to $N - 1$.

The outdegree of a node $i$ is defined as the number of edges emanating from $i$. Similarly, the indegree of node $i$ corresponds to the number of edges pointing towards that node. Nodes with particularly high degree values are called hubs. The concept of node degree can be generalized hierarchically [16-18]. The hierarchical outdegree of a node $i$ is henceforth understood as the number of edges extending from the ring of radius $d$ centered at $i$ to the nodes belonging to the ring of radius $d+1$ centered at that same node, i.e. $R_d(i)$. For example, the hierarchical degrees of node 5 in Figure 1 are $h_0(5) = 5$, $h_1(5) = 6$ and $h_2(5) = 0$. It is also possible to define the hierarchical indegree, but this measurement will not be considered in the present work.

Although the number of hierarchical neighbors and the hierarchical degree are often correlated, they are usually not identical. For instance, there are 4 nodes in $R_2(5)$, but 6 edges extend from $R_1(5)$ to $R_2(5)$, indicating that some nodes in $R_1(5)$ connect to more than one node in $R_2(5)$. As the relative behavior of the number of hierarchical neighbors and hierarchical degree may provide insights about the connectivity of the analyzed network, it is interesting to consider the divergence measurement [18] defined as:

$$D_d(i) = \frac{n_{d+1}(i)}{h_d(i)}$$

For example, we have in Figure 1 that $D_1(5) = n_2(5)/h_1(5) = 4/6 = 2/3$, indicating that there are more edges converging than diverging from ring 1 to ring 2. Because the number of hierarchical neighbors of a generic node $i$ at distance $d = 1$ is always equal to the outdegree of that node, we necessarily have maximum divergence characterized by $D_1(i) = 1$. Note that $0 \leq D_d(i) \leq 1$, with lower values of $D_d(i)$ indicating greater convergence (less divergence).

The hierarchical clustering coefficient [17,18] is another measurement which has been found to be valuable for the characterization of complex networks (e.g. [18,19]). Given a node $i$, its hierarchical clustering coefficient at distance d is defined as the ratio between the number of existing edges in $R_d(i)$ and the maximum possible number of edges between the nodes in that ring, i.e.

$$CC_d(i) = \frac{e_d(i)}{n_d(i)(n_d(i)-1)}$$

where $e_d(i)$ stands for the number of edges inside the ring $R_d(i)$. The hierarchical clustering coefficient therefore provides an interesting means for quantifying the connectivity between nodes at successive distances from the reference node, providing a natural and intuitive extension of the traditional clustering coefficient (e.g. [3,4]). Note that this definition of clustering coefficient takes into account only the outgoing edges of each node. Although an analogue measurement could be defined for the incoming edges, this is not considered in the present paper. The verification of trends (e.g. constant value, monotonic increase or decrease) of $CC_d(i)$ as $d$ increases may indicate that at least a portion of the network connections are organized with respect to the reference node $i$.

3. CONNECTIVITY DATA SETS

Data sets used in this study were identical to those used in [14] and [15]. We examined two large-scale cortical connection matrices, for macaque visual cortex ($N = 30$, $K = 311$; [20]) and cat cortex ($N = 52$, $K = 820$; Scannell et al., 1999). The matrix of macaque visual cortex was modified by eliminating areas PIT, CIT and STP and assigning their connections to {PITd, PITv}, {CITd, CITv} and {STPp, STPa}, respectively, as well as excluding areas MIP and MDP which lacked sufficient connectional information. Following the discussion in [12], the remaining 30 areas can be divided into two functionally distinct clusters, defined by similarities and dissimilarities in their interconnectivity: a parietal and occipito-parietal cluster (V1, V2, P, V3A, MT, V4t, V4, PIP, LIP, VIP, DP, PO, MSTi, MSTd, FST, FEF) and an inferior-temporal and prefrontal cluster (PITv, PITd, CITv, CITd, AITv, AITd, STPa, 7a, TF, TH, VOT, 46). We refer to these two clusters as "dorsal" and "ventral", respectively.

The connection matrix of cat cortex was binarized, and area Hipp (hippocampus) and all thalamo-cortical pathways were excluded. Following [12,22] the remaining 52 cortical areas can be divided into four functionally distinct clusters: visual (17, 18, 19, PLLS, PMLS, ALLS, AMLS, VLS, DLS, 21a, 21b, 20a, 20b, 7, AES, PS), auditory (AI, AII, AAF, P, VPc, EPp, Tem), somatomotor (31, 3b, 1, 2, SII, SIV, 4g, 4, 6l, 6m, 5Am, 5Al, 5Bm, 5Bl, SSAi, SSAo), and frontolimbic (PFCMil, PFCMd, PFCL, Ia, Ig, CGa, CGp, RS, 35, 36, pSb, Sb, Enr). For statistical comparison, visual and auditory clusters are combined into a "posterior" cluster, while somatomotor and frontolimbic clusters are combined into an "anterior" cluster. Additional comparisons are carried out between frontolimbic areas and a set of lower visual and auditory areas (selected based on the ordering reported in [23]) consisting of areas 17, 18, VLS, PS, 19, PMLS, PLLS, AI, AII, AAF, and P, referred to as "sensory".

All statistical comparisons in this paper refer to standard independent-measures t-tests.

The four hierarchical measurements considered in this paper (i.e. hierarchical number of neighbors, hierarchical degree, divergence ratio and hierarchical clustering coefficient) have been from these matrix by using a customized algorithm, written in SCILAB, which first identifies the distances from each reference node to all other nodes and then uses this information in order to calculate the hierarchical degree, divergence ratio and hierarchical clustering coefficient. The identification of the nodes at successive distances from the reference node is performed by using two lists, **L1** and **L2**, the former containing the nodes at the current distance and the latter storing the nodes which can be reached, through a single edge, from the nodes in **L1**. The content of these lists is swapped at each step, until no nodes remain in **L2**.

4. RESULTS AND DISCUSSION

4.1. Macaque Visual Cortex

The polar diagrams in Figure 2 represent a summary (or fingerprint [24]) of hierarchical measures applied to each of the individual areas of macaque visual cortex, including the number of hierarchical neighbors, hierarchical degrees, divergence coefficients and hierarchical clustering coefficients of each considered cortical region as a function of the hierarchical distance $d$. Since the diameter of the network is 3, no region exhibits hierarchical depth higher than 3.

Closer analysis of the hierarchical fingerprint data reveals that individual brain regions show marked differences. For example, an examination of the hierarchical number of nodes allows one to see how much of the network can be accessed from a given reference node at each level of hierarchical depth. Only 6 areas (V2, V4, MT, FST, MSTd, LIP) connect to more than half of the remaining $N-1$ nodes at hierarchical depth 1. All of these areas belong to the dorsal cluster. Only 8 areas connect to less than 90% of the macaque visual cortex up to hierarchical depth 2, and these areas (VOT, CITd, CITv, AITd, AITv, STPp, STPa, 46) all belong to the ventral cluster. Only 8 areas connect to the entire network at hierarchical depth 2, and these areas (V3, VP, V4, MT, TF, MSTd, PIP, LIP) all belong to the dorsal cluster. On average, members of the ventral cluster connect to $7.69 \pm 3.17$ areas at hierarchical depth 1, while members of the dorsal cluster connect to $12.41 \pm 3.73$ areas (p<0.01)

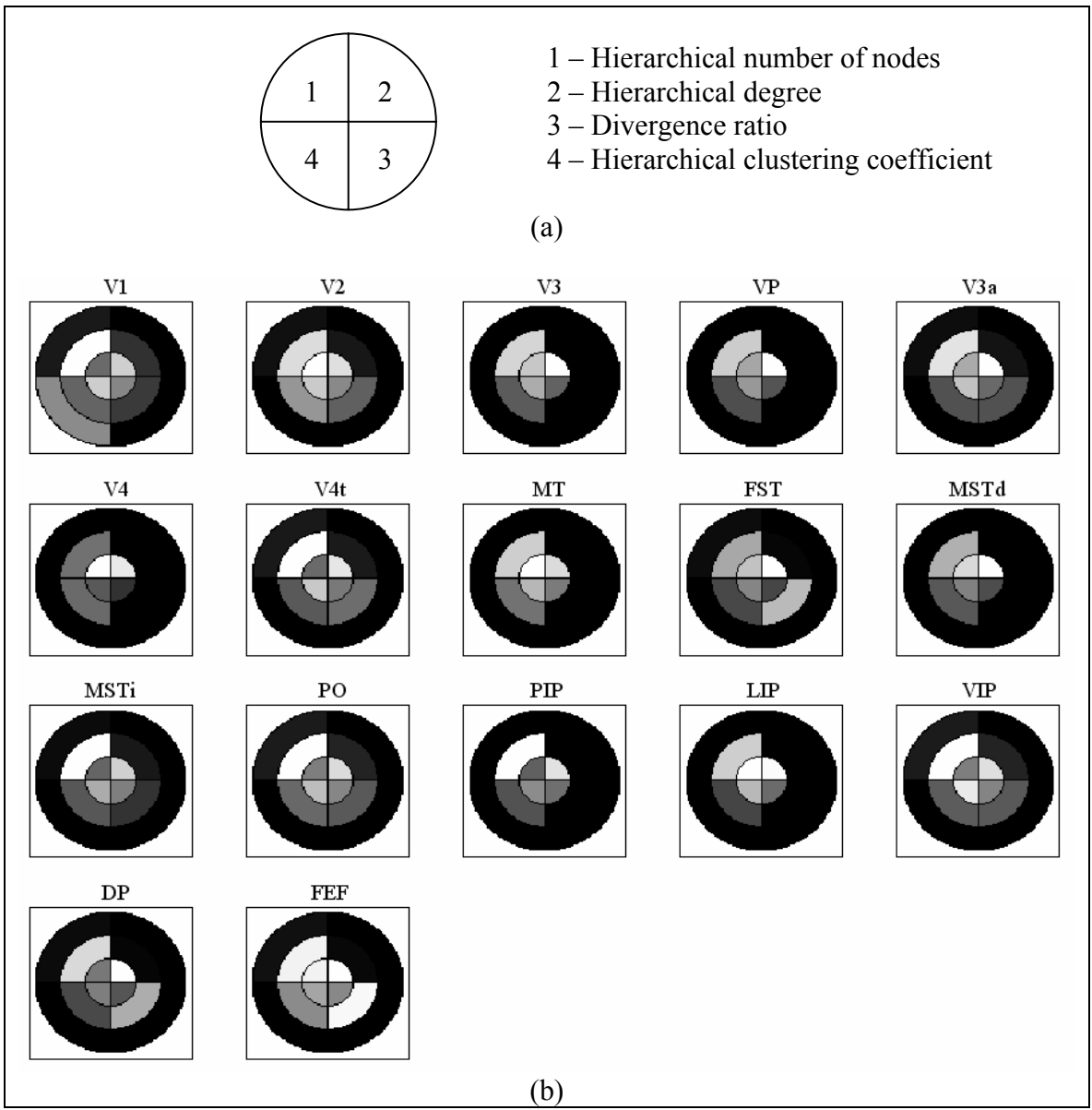

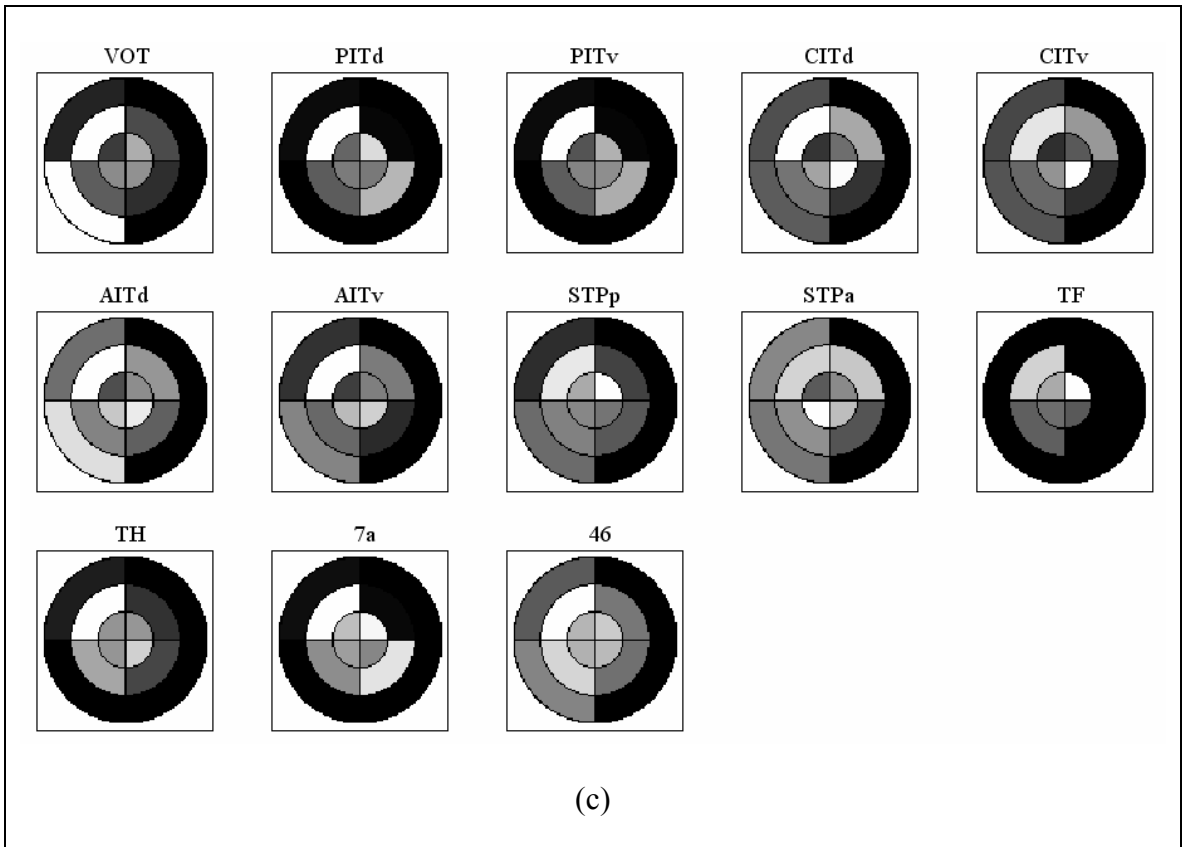

Figure 2 – Polar diagrams showing the number of hierarchical neighbors, hierarchical degrees, divergence ratios and hierarchical clustering coefficients of the considered macaque visual cortical areas, as indicated in the legend (a). The hierarchical depth of all regions is limited to 3, represented by the successive rings, and the gray-level scales are normalized between black (minimum value) and white (maximum value) for each type of hierarchical measure, respectively. The two main clusters correspond to the occipito-parietal or dorsal (b) and inferior-temporal and prefrontal or ventral areas (c).

The hierarchical degree distributions show significant ($p<0.001$) differences for hierarchical depths 1 and 2 (all hierarchical degrees are zero for greater depths). Mean hierarchical degrees for dorsal areas are $59.29 \pm 7.63$ and $3.59 \pm 4.12$, versus $43.54 \pm 13.49$ and $21.46 \pm 17.41$, for depths 1 and 2 respectively. Only four areas exhibit peaks in their hierarchical degree at depth 2 (CITd, CITv, AITd, STPa), all of which are members of the ventral cluster.

The distributions of hierarchical clustering coefficients at depths 2 and 3 show significant ($p<0.001$ and $p<0.01$, respectively) differences between dorsal and ventral areas. Ventral areas are more clustered than dorsal areas (depth 2: $0.42 \pm 0.07$ versus $0.33 \pm 0.05$; depth 3: $0.30 \pm 0.31$ versus $0.03 \pm 0.12$). Notably, no such difference exists at depth 1 ($0.54 \pm 0.11$ versus $0.58 \pm 0.11$). Three areas exhibit greater clustering coefficients at depth 2 than 1, including one dorsal area (V4) and two ventral areas (TH, 46). Two areas have higher clustering coefficients for depth 3 than depth 1, both (VOT, AITd) members of the ventral cluster. Because the hierarchical clustering coefficient is relative to the number of nodes at each depth, this information should be taken into account when interpreting the clustering coefficients. For instance, values of this measurement close to unity that involve only a few nodes are not particularly meaningful. The divergence ratio at depth 1 differs significantly ($p<0.001$) between dorsal ($0.27 \pm 0.06$) and ventral ($0.44 \pm 0.15$) clusters. No such difference was found at depth 2.

Figure 3 shows a scatterplot obtained by considering the hierarchical clustering coefficient for $d = 2$ and the divergence ratio for $d = 1$, displaying differences between the dorsal and ventral clusters of macaque visual cortex.

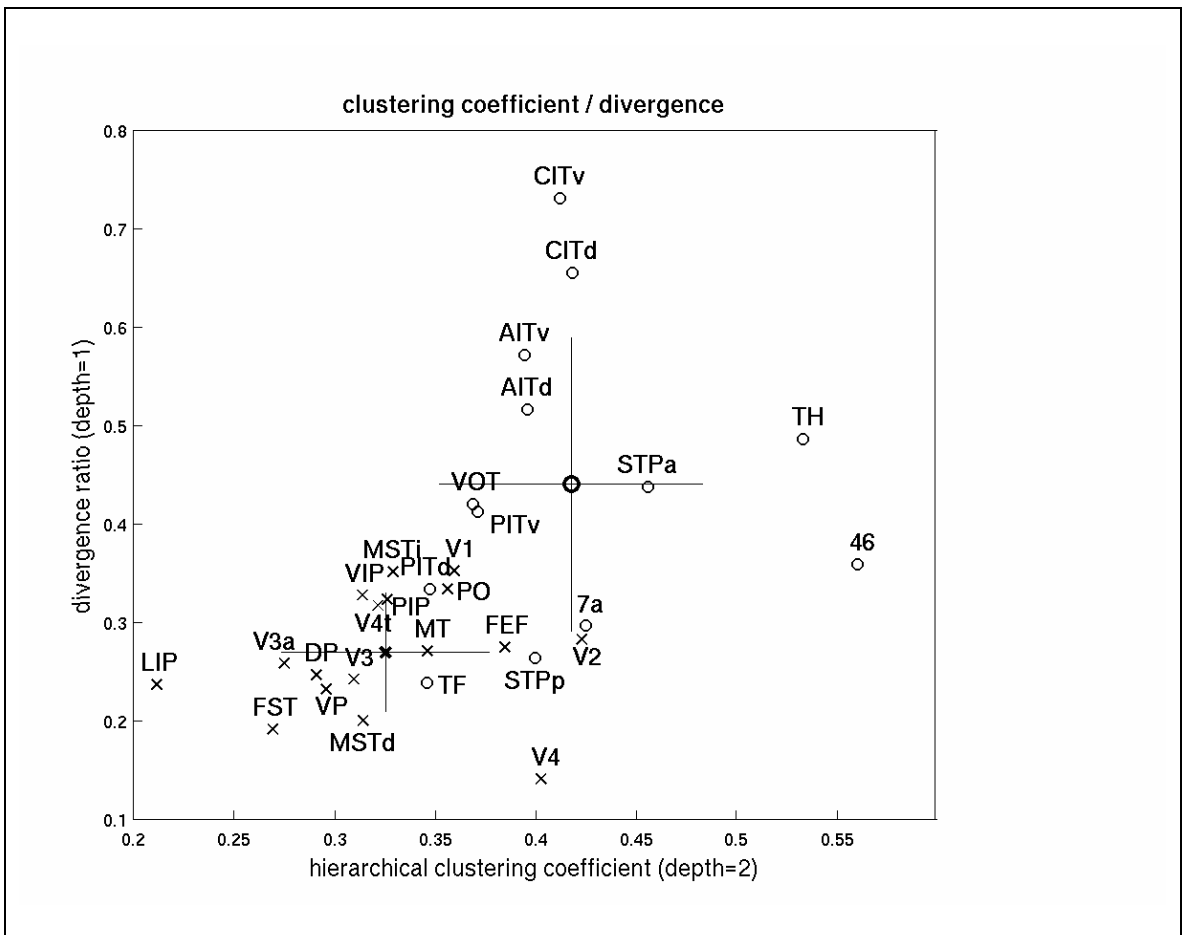

Figure 3 – Scatterplot obtained by considering the hierarchical clustering coefficient for $d = 2$ and the divergence ratio for $d = 1$. Dorsal areas are denoted by 'x', while ventral areas are denoted by 'o'. Fat symbols and errorbars indicate means and standard deviations for dorsal ('x') and ventral ('o') clusters.

To further elucidate the relationships between macaque visual cortical areas in terms of hierarchical connectivity measures, we performed principal components analysis (e.g. [25]) as well as k-means clustering on the data. PCA involves the calculation of the covariance matrix for a set of considered measurements and the projection of those measurements over a subset of eigenvectors of the covariance matrix which are associated with the two largest eigenvalues, resulting in a projection onto those hyperplanes that capture maximal variance. Figure 4 shows the results of PCA applied to hierarchical degree data. Areas belonging to the dorsal and ventral clusters are partially segregated, with areas TH, 46, VOT, AITd, AITv, CITv, CITd, and STPa (all ventral) appearing clearly separated from the remainder of the network. Similar distributions result from PCA on data on hierarchical clustering coefficient and divergence ratio. K-means clustering identifies these same areas for all three hierarchical measures.

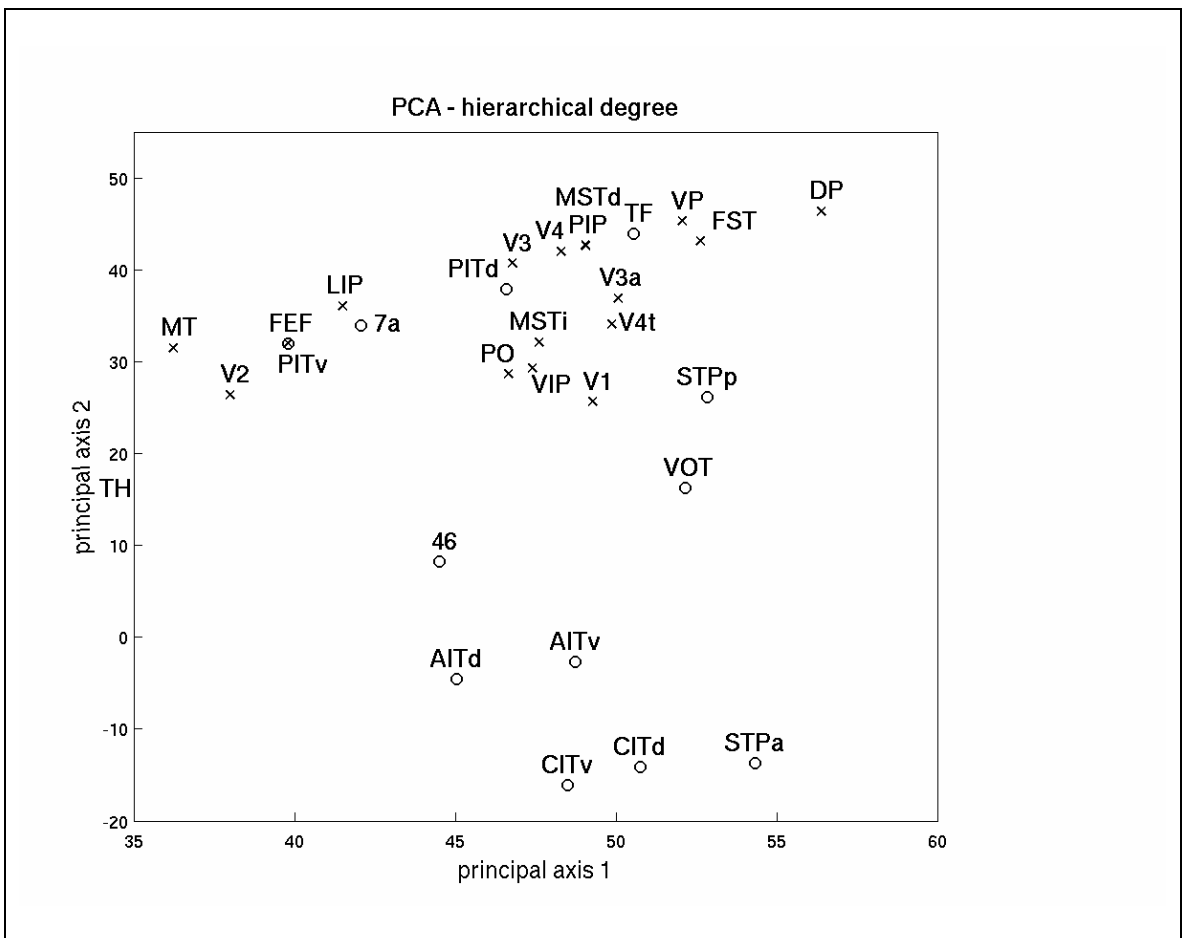

Figure 4 – The projection, through principal component analysis, of the hierarchical degree onto the respective plane of maximum dispersion.

To determine the extent to which the structure revealed by hierarchical measures depends on local connection properties (such as the indegree and outdegree of each node) versus more global patterns, we performed a comparative PCA analysis on hierarchical measures of networks that were randomly rewired (with $N = 30$, and $K = 311$) in such a way as to preserve the degree sequences of individual nodes [28]. We obtained 100 randomized networks which were compared to the real data by considering their principal component projections. Figure 5 illustrates the approach. Most significant differences between macaque brain areas and their randomized counterparts were found for areas TH, 46, AITd, AITv, CITv, and STPa (all ventral).

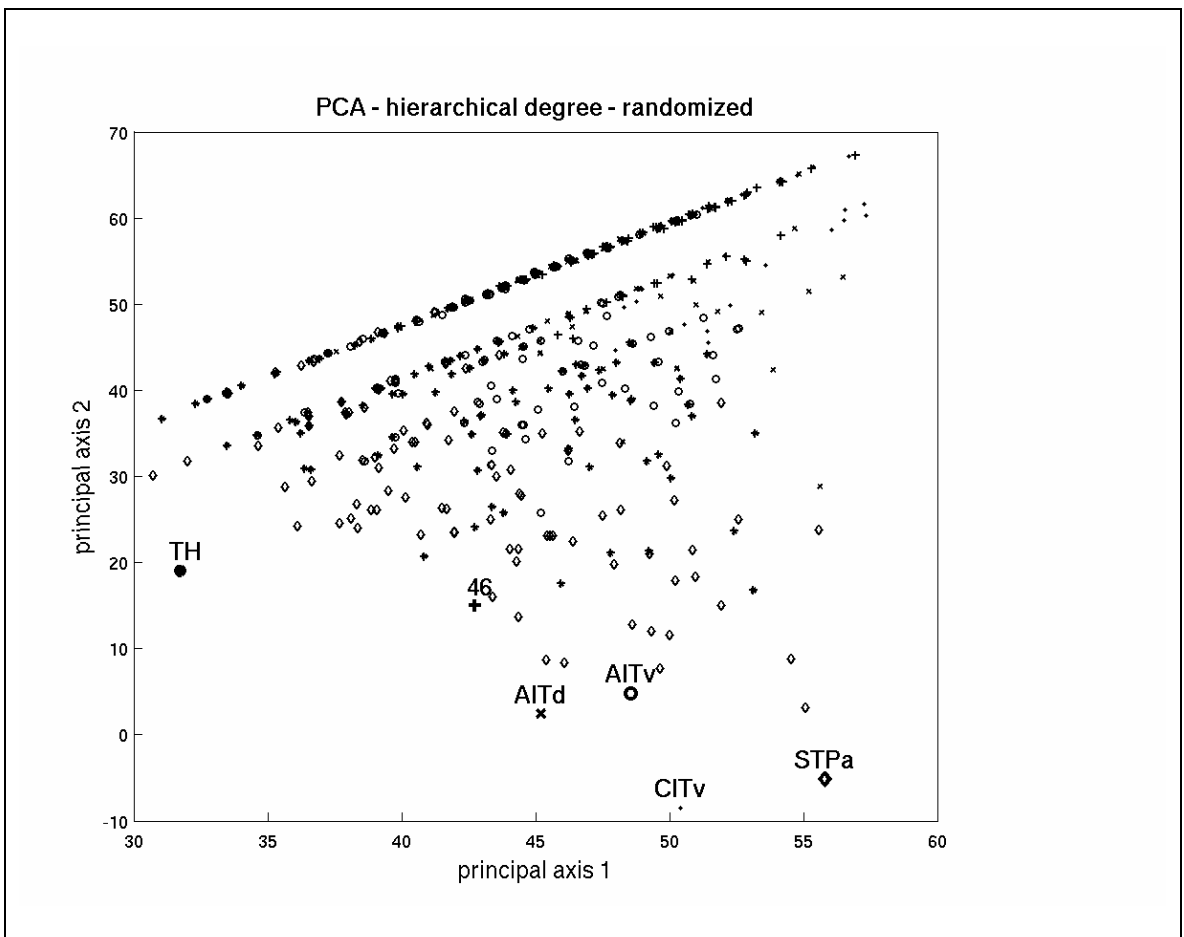

Figure 5 – The hierarchical degree of the macaque visual areas (identified by the letters) and 100 random counterparts projected into the plan of maximum dispersion considering only areas TH, 46, AITd, AITv, CITv, and STPa (all ventral) which showed marked differences between the real and randomized data. The random graphs were obtained so as to preserve the indegrees and outdegrees of each respective area.

4.2. Cat Cortex

We performed hierarchical measures analyses on the connectivity matrix of cat cortical regions, by statistically comparing areas in anterior and posterior clusters (see polar plots in Figure 6), as well as in frontolimbic and sensory clusters (see Section 3 for area memberships). Both comparisons yielded similar results for all four hierarchical measures investigated in this study, and we focus on the comparison between frontolimbic (a subset of anterior) and sensory (a subset of posterior) brain regions.

At depth 1, sensory regions contacted significantly ($p<0.01$) fewer regions than frontolimbic regions ($10.09 \pm 3.42$ versus $18.84 \pm 9.41$, respectively), while this relationship is reversed for depth 3 ($10.91 \pm 10.26$ versus $3.31 \pm 5.72$, $p < 0.05$). This pattern suggests that frontolimbic regions maintain more widespread connections and access the complete network in fewer steps. Data for hierarchical degree provides further support for this hypothesis, with significantly higher degree values for frontolimbic versus sensory regions at depth 1 ($149.00 \pm 44.81$ versus $87.82 \pm 37.54$, $p < 0.01$) and a reversal at depth 2 ($26.92 \pm 42.59$ versus $74.82 \pm 46.48$, $p< 0.05$).

While the hierarchical clustering coefficient did not exhibit significant differences at depth 1, frontolimbic areas showed a lower clustering coefficient at depth 2 compared to sensory areas ($0.29 \pm 0.05$ versus $0.41 \pm 0.13$, $p< 0.01$). The divergence ratios for depths 1 and 2 were significantly lower for frontolimbic areas versus sensory areas (depth 1: $0.24 \pm 0.16$ versus $0.36 \pm 0.08$, $p< 0.05$; depth 2: $0.05 \pm 0.06$ versus $0.13 \pm 0.06$, $p<0.01$). These two measures indicate that frontolimbic areas are less segregated and more divergent in terms of their connection structure, compared to sensory areas.

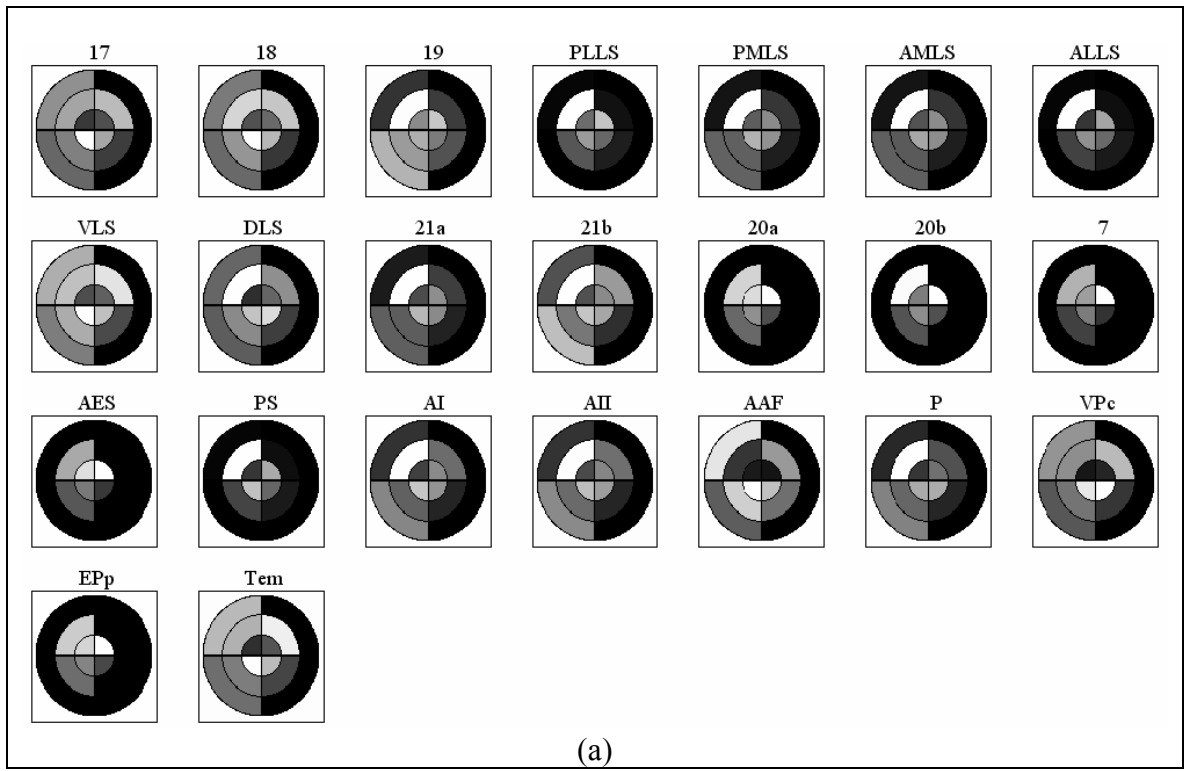

(a)

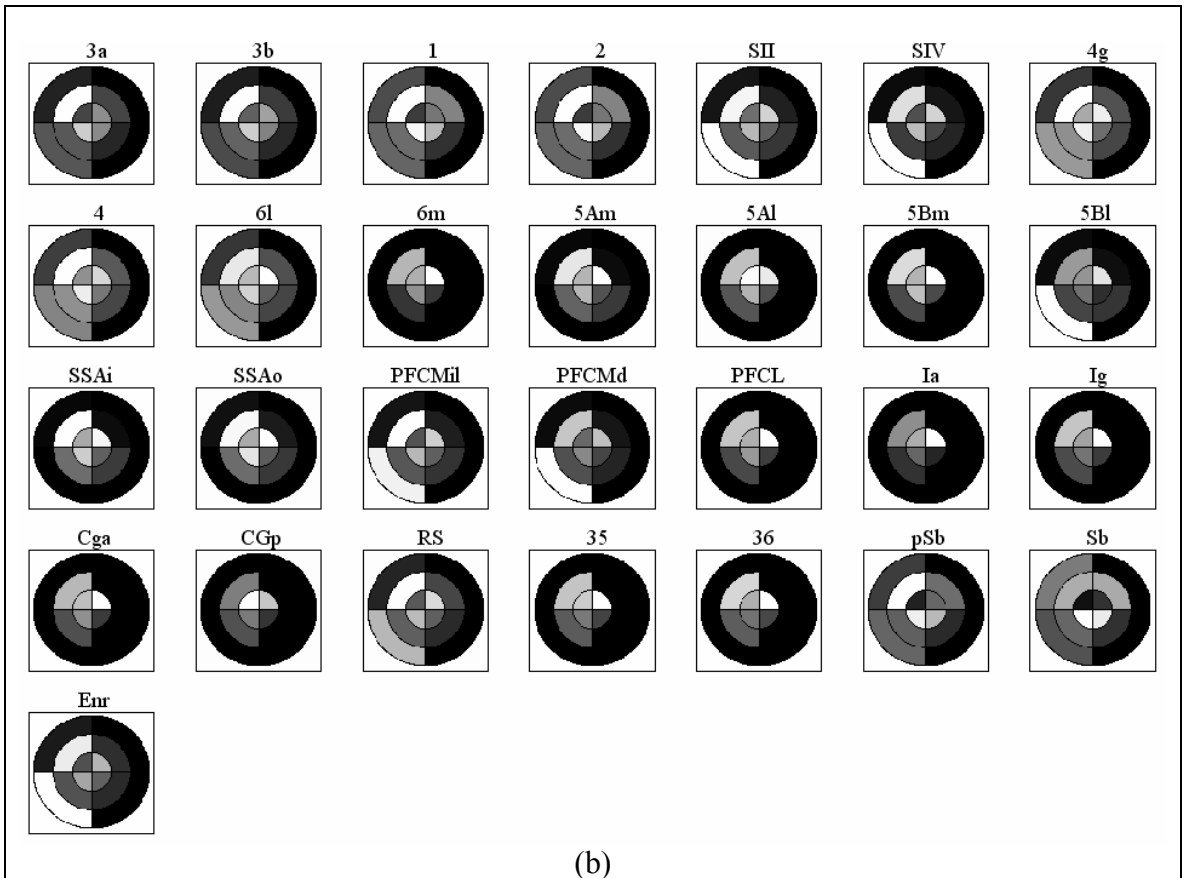

(b)

Figure 6 – Polar diagrams (legend as in Figure 2) expressing the four hierarchical measures obtained for the cat data. The regions are organized into two main groups, corresponding to the posterior (a) and anterior (b) clusters.

## 5. CONCLUDING REMARKS

Our study shows that the use of hierarchical measures of connectivity can reveal significant differences in the way a given reference node accesses and interacts with the rest of the surrounding network. Note that our goal was not to identify unique hierarchical arrangements of brain regions, in terms of processing stages of streams, an approach taken in earlier work [20, 12, 23]. Instead we apply hierarchical measures to each brain region, thus placing it as a reference node at the center of the network, and we examine how the efferent connections of this node extend to successive levels of the surrounding network.

In generating hierarchical network measures for each brain region we extend the concept of "connectivity fingerprints" [27], "network participation indices" [24], and "motif fingerprints" [15], which attempt to assess regional contributions to global network architecture. Our analysis is consistent with earlier studies that had attributed differences in terms of such contributions between dorsal and ventral processing streams in macaque visual cortex. We reveal that the ventral stream consists of areas that are more tightly clustered at hierarchical depths greater than 1 and exhibit less divergence, when compared to areas of the dorsal stream. This may point to important functional differences between these two visual cortical subdivisions. Extending the analysis to cat cortex we find that clusters of sensory areas are less divergent (i.e. functionally more segregated) while regions of the frontolimbic complex, containing many polysensory and multimodal neurons, are more divergent and less highly clustered at depths greater than 1. This provides objective, connectivity-based evidence for their diverse effects on broad regions of cortex, subserving their functional integration.

One major application of hierarchical measures, highlighted in this paper, is in identifying principles of structural network organization. In addition, we point to several important functional or dynamic connotations of the present analysis. Since hierarchical measures are computed based on the efferent connectivity of a reference node, they essentially capture the spatio-temporal spread of information away from that node as it becomes activated. Such activity propagation can be experimentally assessed [28] and may be a major ingredient in the spreading of epileptic seizures originating in a cortical locale. We note that hierarchical measures can also be calculated on the basis of afferent connections terminating on a reference node, in which case functional impact of the network onto the node can be assessed.

There are many future applications of hierarchical measures, including the study of additional large-scale connection matrices such as the one of whole macaque cortex [29], or of cat cortex including cortico-thalamic pathways [21]. Analyses carried out in other mammalian species may provide important information on the evolutionary progression of brain connectivity. Finally, additional hierarchical measures may be devised that allow further insights into the organization of complex networks.

Acknowledgments

Luciano da F. Costa is grateful to FAPESP (proc.99/12765-2), CNPq (proc. 308231/03-1) and Human Frontier Science Program (proc. RGP39/2002).

BIBLIOGRAPHY

[1] – G. Tononi, G.M. Edelman, O. Sporns, *Trends Cogn. Sci.* **2**: 474-484, 1998.

[2] – K.J. Friston, *Annu. Rev. Neurosci.* **25**: 221-250, 2002.

[3] - R. Albert and A.-L. Barabási, *Rev. Mod. Phys.*, **74**:48-98, 2002.

[4] – M. E. J. Newman, *SIAM Review*, **45**(2): 167-256, 2003.

[5] – S. N. Dorogovtsev and J. F. F. Mendes, *Adv. in Phys.*, **51**: 1079-1187, 2002.

[6] – L. da F. Costa, F. A. Rodrigues, G. Travieso and P. R. Villas-Boas, cond-mat/0505185, May 2005.

[7] – G. Buzsaki, C. Geisler, D.A. Henze, and X.-J. Wang, *Trends Neurosci.* **27**(4): 183-193, 2004.

[8] – L. da F. Costa, q-bio.MN/0503041, Mar 2005.

[9] – B. Amirikian, *PLoS Comp. Biol.* **1**(1): 74-85, 2005.

[10] – D. Stauffer, A. Aharony, L. da F. Costa, and J. Adler, *Eur. Phys. J. B*, **32**: 395-399, 2003.

[11] – L. da F. Costa and D. Stauffer, *Phys. A*, **330**(1-2): 37-45, 2003.

[12] – C.C. Hilgetag, G.A.P.C. Burns, M.A. O'Neill, J.W. Scannell, M.P. Young, *Philos. Trans. R. Soc. Lond. B. Biol. Sci.* **355**: 91-110, 2000.

[13] – O. Sporns, G. Tononi, G.M. Edelman, *Cereb. Cortex* **10**: 127-141, 2000.

[14] – O. Sporns and J. D. Zwi, *Neuroinformatics*, **2**(2): 145-162, 2004.

[15] – O. Sporns and R. Kötter, *PLoS Biol.* **2**(11): 1910-1918, 2004.

[16] - L. da F. Costa, Phys. Rev. Lett. 93: 098702, 2004.

[17] - L. da F. Costa, cond-mat/0408076, Aug. 2004.

[18] - L. da F. Costa, cond-mat/0412761, Dec 2004.

[19] - L. da F. Costa, cond-mat/0501010, Jan. 2005.

[20] – D.J. Felleman and D.C. Van Essen, *Cereb. Cortex* **1**: 1-47, 1991.

[21] – J.W. Scannell, G.A.P.C. Burns, C.C. Hilgetag, M.A. O'Neill, M.P. Young, *Cereb. Cortex* **9**: 277-299, 1999.

[22] – C.C. Hilgetag and M. Kaiser, *Neuroinformatics* **2**(3): 353-360, 2004.

[23 ] – J.W. Scannell, C. Blakemore, M.P. Young, J. Neurosci. 15: 1463-1483, 1995.

[24] – R. Kötter and K.E. Stephan, *Neural Netw*. **16**: 1261-1275, 2003.

[25] - L. da F. Costa and R. M. Cesar Jr., *Shape Analysis and Classification*, CRC Press, 2001.

[26] – R. Milo, S. Shen-Orr, S. Itzkovitz, N. Kashtan, D. Chklovskii, U. Alon, *Science* **298**: 824-827, 2002.

[27] – R.E. Passingham, K.E. Stephan, R. Kötter, *Nat. Rev. Neurosci.* **3**: 606-616.

[28] – R. Kötter and F.T. Sommer, Phil. Trans. R. Soc. Lond. B 355: 127-134, 2000.

[29] - M.P. Young, Proc. R. Soc. Lond. B. Bio. Sci. 252: 13-18, 1993.